\documentclass[10pt,twocolumn]{paper}
\usepackage{epsfig, graphicx}
\usepackage[english]{babel}
\usepackage[margin=2.5cm]{geometry}

\title{\center \rm \bf Thermotropic Phase Transition in an Adsorbed Melissic Acid Film at the n-Hexane -- Water Interface}

\author{\small \rm Aleksey M. Tikhonov\/\thanks{tikhonov@kapitza.ras.ru}\\ \small 
Kapitza Institute for Physical Problems, Russian Academy of Sciences, Moscow, Russia}

\begin{document}
\maketitle
%\centerline{\today}

\abstract{ \it A reversible thermotropic phase transition in an adsorption melissic acid film at the interface between n-hexane and an aqueous solution of potassium hydroxide (pH\,$\approx 10$) is investigated by X-ray reflectometry and diffuse scattering using synchrotron radiation. The experimental data indicate that the interface "freezing" transition is accompanied not only by the crystallization of the Gibbs monolayer but also by the formation of a planar smectic structure in the ~300-\AA-thick adsorption film; this structure is formed by ~50-\AA-thick layers.}

\vspace{0.25in}
%\large

An adsorption film at an oil -- water interface can be viewed as a two-dimensional thermodynamic system
characterized by a set of parameters $(p,T,c)$. This system can be isotropic or anisotropic even in the case of isotropic bulk phases [1-4]. It was reported earlier that melissic acid (C$_{30}$-acid) is adsorbed at the interface between n-hexane and an aqueous solution of sulfuric acid (pH\,$\approx 2$) as a protonated multilayer in which there exist three thermotropic mesophases [5]. On one hand, an increase in the temperature $T$(at a pressure of $p=1$\,atm) leads to a solid -- liquid phase transition
in the monolayer immediately located at the interface (Gibbs monolayer), the transition temperature $T_c$
being determined by the concentration of the surfactant in the bulk phase of n-hexane, which serves as a
reservoir for the former [6]. On the other hand, upon a decrease in the temperature, the two-dimensional
crystallization phase transition at the interface is preceded by a transition at $T^*>T_c$ to multilayer adsorption. At $T>T^*$, there is only a liquid Gibbs monolayer with a thickness of $(36 \pm 2 )$\,\AA{} at the interface. Data obtained for this system by X-ray scattering and reflectometry can be qualitatively understood in the framework of a three-layer model of the interface, sketched in Fig. 1: for $T<T^*$, in addition to the Gibbs monolayer (layers 1, 2), there is a "thick" uniform layer of the high-molecular-weight alkane liquid (layer 3). Here, we investigate a system with a high pH level in the aqueous phase (pH$\approx 10$) and show that, at $T = T_c$, apart from the solidification of the partially ionized Gibbs monolayer, there occurs a transition in thick layer 3 leading to the reversible formation of a 300-\AA-thick planar smectic structure consisting of 50-\AA-thick layers; i.e., there exists a fourth surface mesophase.

\begin{figure}
\hspace{0.3in}
\epsfig{file=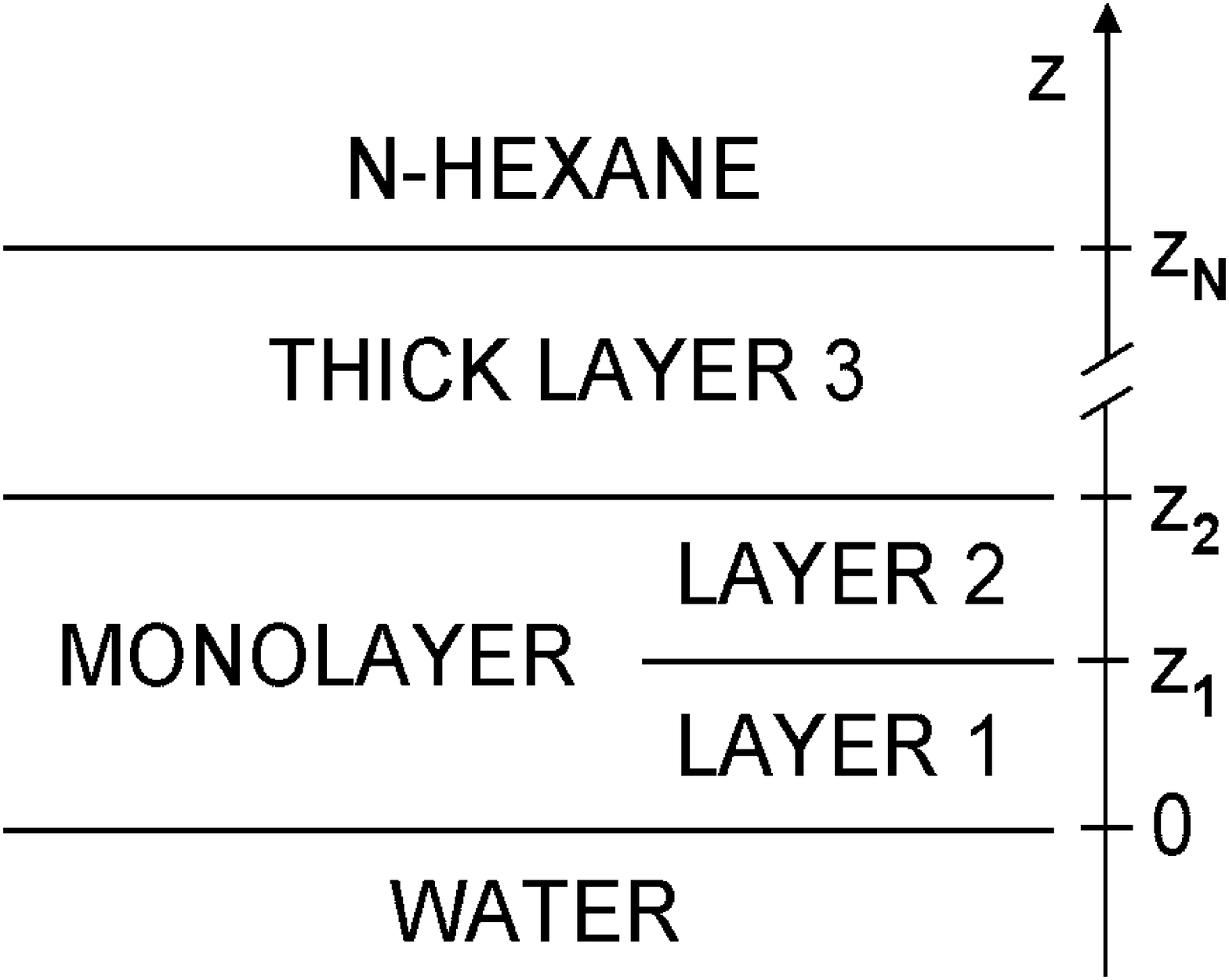, width=0.35\textwidth}

\small {\bf Figure 1}. Three-layer model of the adsorption film of melissic acid C$_{30}$H$_{60}$O$_2$ at the n-hexane -- water interface.
\end{figure}

Samples with a $75\times150$-mm macroscopically flat n-hexane -- water interface were prepared and investigated
in a air-tight stainless-steel cell with polyester windows transparent to X-rays according to the technique
described in [7, 8]. In X-ray scattering measurements, the cell temperature was controlled by a two-stage
thermostat. Saturated hydrocarbon n-hexane (C$_6$H$_{14}$) with the boiling temperature of $T_b\approx 342$\,K and the density of about 0.65 g/cm$^3$ at 298\,K was purified by repetitive filtering in a chromatographic column [9]. The bulk concentration of the C$_{30}$-acid in n-hexane in the systems under study was $c\approx 0.3$\,mmol/kg $\approx 2\cdot10^{-5}$). The $\approx 41$-\AA-long linear chain amphiphilic molecule of the C$_{30}$H$_{60}$O$_2$ acid has a hydrophilic head group (-COOH) and a hydrophobic hydrocarbon tail group (-C$_{29}$H$_{59}$). The volume of the oil phase in the cell was $\sim 100$\,mL, and the amount of melissic acid dissolved in this volume is sufficient to coat the interface with $> 10^2$ acid monolayers. The bottom bulk phase, in which the C$_{30}$-acid is almost insoluble, was formed by solutions of KOH in deionized water (Barnstead, NanoPureUV) with pH\,$\approx 10$.

\begin{figure}
\hspace{0.1in}
\epsfig{file=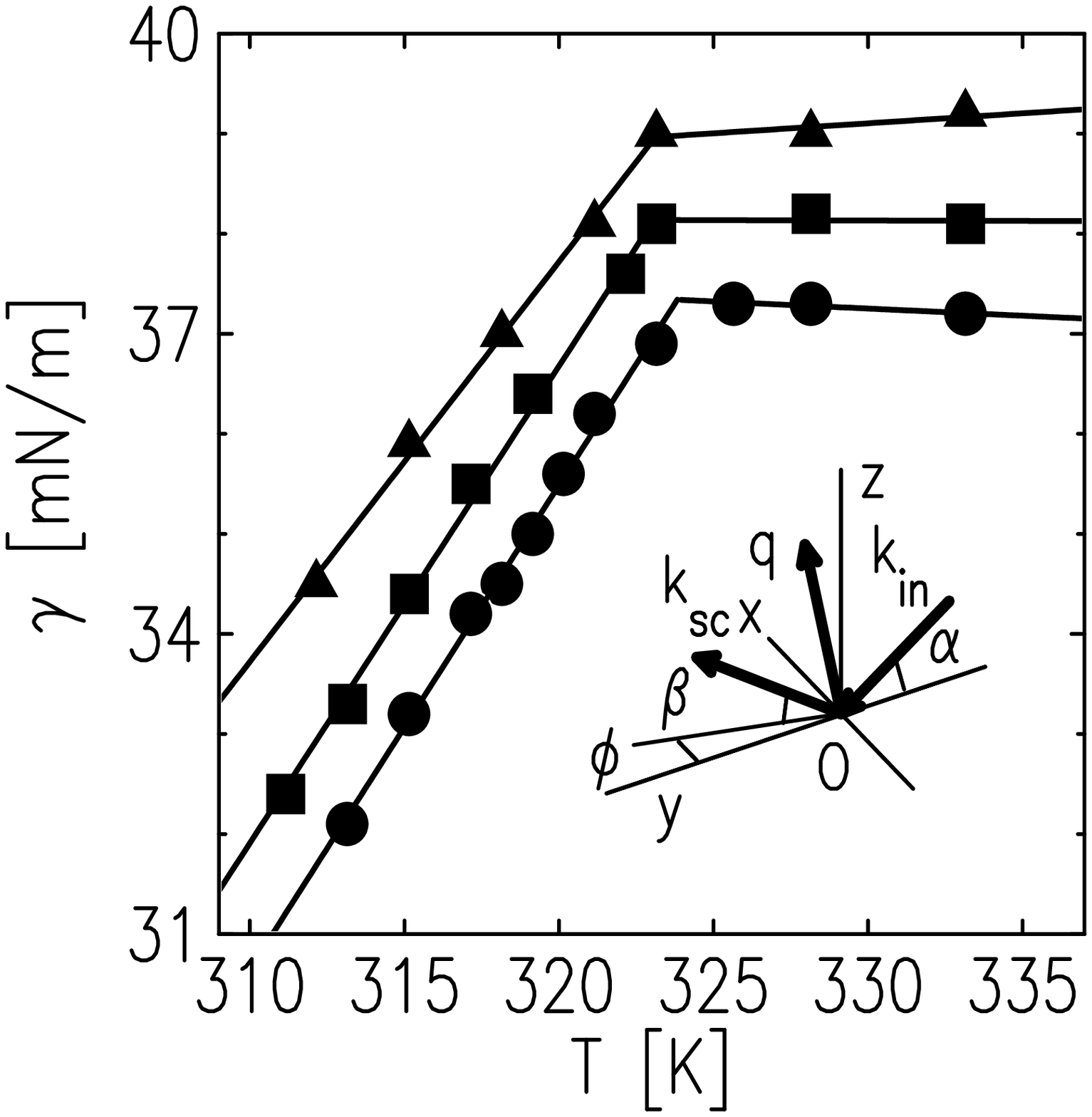, width=0.37\textwidth}

\small {\bf Figure 2}. Temperature dependences of the interfacial tension at the n-hexane~-- water interface for different concentrations of KOH in water with pH = 9.7, 10.1, and 10.5 (circles, squares, and triangles, respectively). Lines show the approximation of $\gamma(T)$ by linear functions. The inflection point corresponds to $T_c\approx323$K. Inset: the scattering kinematics is described in the reference frame with the $xy$
plane coinciding with the water–monolayer interface, the $Ox$ axis perpendicular to the beam direction, and the $Oz$ axis normal to the surface and oriented opposite to the gravity force.
\end{figure}

Figure 2 shows the temperature dependences of the interfacial tension $\gamma(T)$ measured by the Wilhelmy
plate method for systems with pH = 9.7, 10.1, and 10.5 (circles, squares, and triangles, respectively) [10]. The dependences feature an abrupt bend at the phase transition temperature $T_c \approx 323$\,Ê. The lines show the linear least-squares fits for $\gamma(T)$. A change in the slope of $\gamma(T)$ is related to the change in the surface enthalpy upon the transition: $\Delta H = - T_c\Delta(\partial \gamma/\partial T)_{p,c}$  $=0.13\pm 0.02$\,J/m$^2$.

The measurements of the reflection coefficient $R$ and the intensity of diffuse surface scattering $I_n$ of X rays at the n-hexane -- water interface were carried out using synchrotron radiation at the X19C beamline of
the National Synchrotron Light Source (NSLS, Brookhaven National Laboratory, United States) [11].
The intensity $I_0$ of the incident monochromatic beam of photons at a wavelength of $\lambda=0.825 \pm 0.002$\,\AA{} was $\sim 10^{10}$\,photon/s.

Let {\bf k}$_{\rm in}$ and {\bf k}$_{\rm sc}$ be the wave vectors with amplitude $k_0= 2\pi/\lambda$
of the incident and scattered beams, respectively. It is convenient to introduce a reference
frame with the origin $O$ at the center of the irradiated region, the plane $xy$ coinciding with the water
boundary, the $Ox$ axis perpendicular to the beam direction, and the $Oz$ axis normal to the surface and
oriented opposite to the gravitational force (see Fig. 2, inset). In experiment, the grazing angle in the
$yz$ plane is $\alpha << 1$ and the scattering angle is $\beta << 1$, while the angle in the $xy$ plane between the direction of the incident beam and the direction of scattering is $\phi \approx 0$. The scattering vector {\bf q = k$_{\rm in}$ {\rm -} k$_{\rm sc}$} has the components $q_x=k_0\cos\beta\sin\phi$$\approx k_0\phi$ and $q_y=k_0(\cos\beta\cos\phi-\cos\alpha)$ $\approx k_0(\alpha^2-\beta^2)/2$ in the interface plane and the component $q_z=k_0(\sin\alpha+\sin\beta)$$\approx k_0(\alpha+\beta)$ perpendicular to this plane.

\begin{figure}
\hspace{0.1in}
\epsfig{file=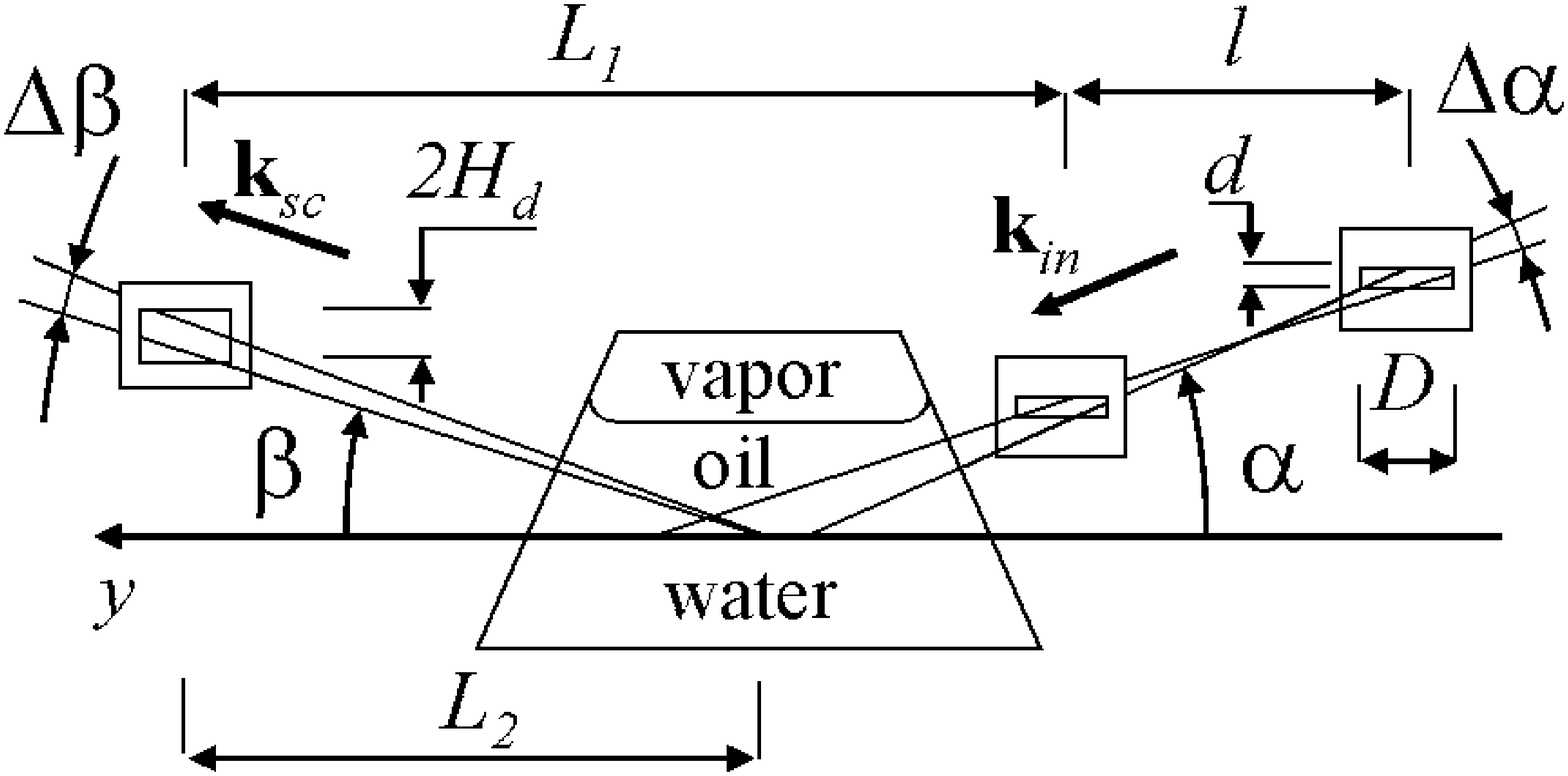, width=0.4\textwidth}

\small {\bf Figure 3}. Parameters of the optical layout.
\end{figure}

The angular divergence $\Delta\alpha=d/l\approx 10^{-4}$\,rad of the incident beam was controlled by a pair of collimating slits with a vertical size of $d=0.05$\,mm spaced by $l\approx 60$\,cm from each other (see Fig. 3). The distance from the collimating slit in front of the sample to the
detector was $L_1\approx 90$\,cm. The size of all slits in the horizontal plane was $D\approx 10$\,mm, which is considerably greater than the horizontal width of the incident beam ($\sim 2$\,mm). In the measurements of the reflection coefficient $R(q_z)$, the collimator slits were gradually
opened with an increase in the angle $\alpha$ to reach the maximum size of $d=0.4$\,mm for $\alpha > 10^{-2}$\,rad. The angular resolution of the point detector in the horizontal
plane was $\Delta \phi = D/L_1$$\approx 10^{-2}$\,rad. The angular resolution in the plane of incidence $\Delta\beta=2H_d/L_2$$\approx 3\cdot10^{-4}$\,rad in the measurements of diffuse scattering intensity is set by a slit of a width $2H_d = 0.2$\,mm placed in front of the detector and at a distance of $L_2\approx 70$\,cm from the center of the sample. In the measurements of the reflection coefficient, $2H_d = 1.6$\,mm.

Figure 4 shows the dependence $R(q_z)$ for the n-hexane~-- water interface at different temperatures
below and above the phase transition. For $q_z < q_c \approx 0.01$\,\AA$^{-1}$, the incident beam experiences total external reflection and $R\approx 1$. The total external
reflection angle $\alpha_c$ ($q_c=2k_0\sin\alpha_c$) is determined by the difference $\Delta\rho\approx0.11$\,{\it e$^-$/}{\AA}$^3$ between the bulk electron
densities in n-hexane and water ($\rho_h\approx0.22$\,{\it e$^-$/}{\AA}$^3$
and $\rho_w\approx0.33$\,{\it e$^-$/}{\AA}$^3$, respectively) and equals $\alpha_c =\lambda\sqrt{r_e\Delta\rho/\pi}\approx 10^{-3}$\,rad (here, $r_e =2.814\cdot10^{-5}$\,\AA is the classical electron radius).

For $T<T_c$, the experimental dependences $R(q_z)$ exhibit a narrow feature ($\delta q_z\approx 0.02$\,\AA$^{-1}$) around $q_z \approx 0.25$\,\AA$^{-1}$. This feature manifests itself as a peak in the plot of the reflection coefficient normalized by the Fresnel function $R_F(q_z)=$ $(q_z-[q_z^2-q_c^2]^{1/2})^2/(q_z+[q_z^2-q_c^2]^{1/2})^2$ (see Fig. 5). With an increase in $T$ the temperature near $T_c$ ($\Delta T\approx 0.5$\,Ê), the shape of the dependence $R(q_z)$ changes and the magnitude of this peak drops abruptly, which gives evidence of the modification of the structure of the absorption film.

The experimental data for the normalized intensity of surface scattering $I_n (\beta) \equiv (I(\beta)-I_b)/I_0$ measured at $\alpha \approx 3.3 \cdot 10^{-3}$\,rad in the temperature range from
319 to 330\,K are shown by circles in Fig. 6. Here, $I(\beta)$ is the number of photons specularly reflected and diffusely scattered by the irradiated region with an area of $A_0\approx 30$\,mm$^2$ at the center of the interface in the $\beta$ direction, $I_b$ is the number of photons scattered in the bulk of n-hexane on their way to the interface, and $I_0$ is the normalization constant proportional to the intensity of the incident beam; the normalization condition is $I_n(\alpha)\equiv 1$). The method to determine $I_b(\beta)$ is described in detail in~[5].

\begin{figure}
\hspace{0.1in}
\epsfig{file=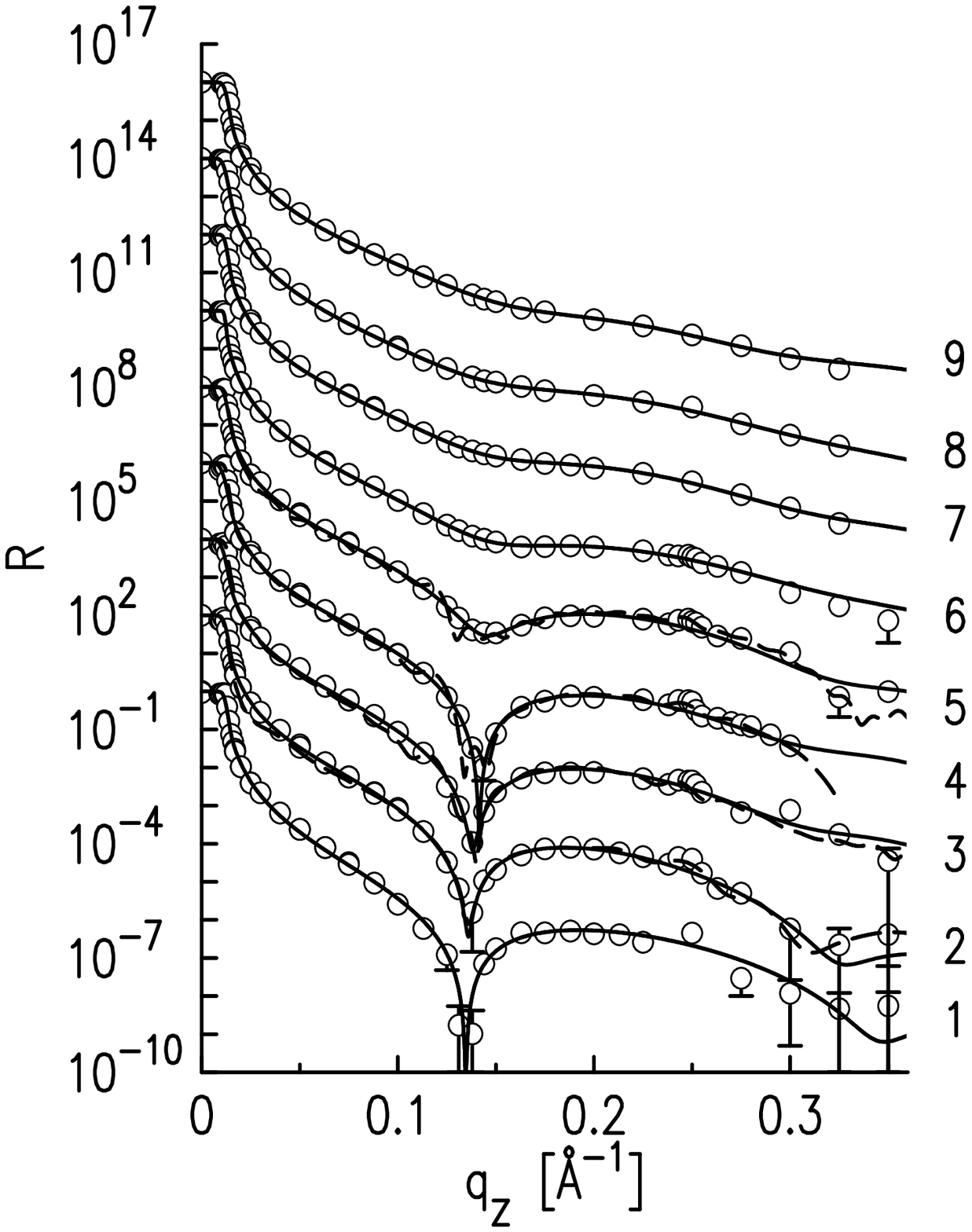, width=0.39\textwidth}

\small {\bf Figure 4}. Reflection coefficient $R$ at the n-hexane -- water interface at temperatures $T$ = (1) 319.3, (2) 320.1, (3) 320.6, (4) 322.1, (5) 323.1, (6) 323.5, (7) 324, (8) 326, and (9) 333.4\,K. Solid lines show the results of calculations for the monolayer model (see Eq. (6)). Dashed lines for $T<T_c$ show the results of calculations in the model with the full width of the surface structure of $\sim 340$\,\AA ($N=8$) and $W\approx 50$\,\AA{} (see Eq. (8)).
\end{figure}

The vertical resolution of the detector $\Delta\beta$ sets a long-wavelength limit of $2\pi/q_y \sim 10^{-5}$\,m on the in-plane lengths involved in scattering. The largest value of $\beta \sim 1.2\cdot 10^{-2}$\,ðàä ($\approx 0.7^\circ$) for which the surface and bulk contributions to the scattering intensity can still be separated from each other sets the short-wavelength limit at $\sim 10^{-6}$\,m.

The most intense peak in Fig. 6 corresponds to specular reflection at $\beta = \alpha$, and the peak in the diffuse background at $\beta \to 0$ corresponds to the total external reflection angle $\alpha_c \approx 10^{-3}$\,rad ($\approx 0.05^\circ$) [12]. For $T<T_c$, on the right shoulder of the specular
reflection peak, there is a smaller peak seen against the diffuse background at $\beta \approx 7\cdot 10^{-3}$. This peak disappears as the temperature $T$ is increased in a small vicinity of $T_c$ ($\Delta T\approx 0.5$\,Ê).

The experimental data were interpreted in terms of qualitative models describing the interface as an ideal
layered structure. Then, in the distorted wave Born approximation, the intensity of surface scattering of a
monochromatic beam of photons can be expressed as the sum of the intensities of diffuse scattering $I_{\rm diff}$ and specular reflection $I_{\rm spec}$ [13, 14]. Thus, the normalized intensity is $I_n \propto I_{\rm diff} + I_{\rm spec}$, where the proportionality coefficient is determined by the normalization condition $I_n(\alpha)\equiv 1$.

Taking into account only the nonspecular scattering of photons by thermal fluctuations of the liquid
surface (capillary waves), we have in the first approximation [5, 15-18]
\begin{equation}
\begin{array}{l}
I_{\rm diff} \approx \frac{\displaystyle \lambda q^4_c }{\displaystyle 512\pi^2}
\frac{\displaystyle k_B T}{\displaystyle \Delta\alpha\gamma}\times
\\ \\
\times
\displaystyle\int\limits_{\alpha-\Delta\alpha/2}^{\alpha+\Delta\alpha/2}
\int\limits_{\beta-\Delta\beta/2}^{\beta+\Delta\beta/2}
\frac{\displaystyle |T(\alpha)|^2|T(\beta)|^2|\Phi(\sqrt{q_zq_z^t})|^2
}{\displaystyle \alpha\sqrt{q^2_y+g\Delta\rho_m/\gamma}} d\beta d\alpha,
\end{array}
\end{equation}
where $q_z^t \approx k_0\left(\sqrt{\alpha^2-\alpha^2_c}+\sqrt{\beta^2-\alpha^2_c}\right)$ is the $z$-component of the scattering vector in the lower phase, $k_B$ is the Boltzmann constant, $g$ is the acceleration of gravity, $\gamma$ is the interfacial tension, $\Delta\rho_m\approx 0.34$\,g/cm3 is
the difference between the densities of water and n-hexane, $\Phi(q)$ is the structure factor of the interface, and $T(\theta)=2\theta/[\theta + \sqrt{\theta^2 - \alpha^2_c}]$ is the Fresnel amplitude
transmission coefficient for a wave polarized in the plane of the interface.

The intensity of specular reflection is given by the expression
\begin{equation}
I_{\rm spec}=f(\alpha, \beta)R(\alpha),
\end{equation}
where the reflection coefficient
\begin{equation}
R(\alpha)=\left|\frac{\displaystyle q_z-q_z^t}{\displaystyle q_z+q_z^t}\right|^2\left|\Phi(\sqrt{q_zq_z^t})\right|^2,
\end{equation}
is calculated under the condition $\alpha \equiv \beta$.

The instrument angular response function $f(\alpha, \beta)$ for a beam with a Gaussian distribution of intensity in the plane of incidence is [18]
\begin{equation}
f(\alpha, \beta)=\frac{1}{2}\left[{\rm erf}\left(\frac{H+H_d}{\sqrt{2}L_1\Delta\alpha}\right) -
{\rm erf}\left(\frac{H-H_d}{\sqrt{2}L_1\Delta\alpha} \right)\right],
\end{equation}
where $H=(\beta - \alpha)L_2$ and ${\rm erf}(x)=(2/\sqrt{\pi})\int_{0}^{x}{e^{-y^2}dy}$ is the error function.

In this model approach, the interpretation of the data is reduced to the finding of the function $\Phi(q)$,
which can generally be written as
\begin{equation}
\Phi(q)=\frac{1}{\Delta\rho}\int^{+\infty}_{-\infty}\left\langle\frac{d\rho(z)}{dz}\right\rangle e^{iqz} dz,
\end{equation}
where $\rho(z)$ is the distribution of the electron density
along the $Oz$ axis averaged over the irradiated area $A_0$.

In the qualitative two-layer model suggested in [6] for the parametrization of the phases of the Gibbs
monolayer of melissic acid, the structure factor has the form
\begin{equation}
\Phi_m(q) = \frac{e^{-\sigma_R^2q^2/2}}{\Delta\rho}\sum_{j=0}^{2}{(\rho_{j+1}-\rho_j)e^{-iqz_j}},
\end{equation}
where $z_0=0$, $\rho_0=\rho_w$, and $\rho_3 = \rho_h$. In the solid phase
of the monolayer, the electron densities are $\rho_1 \approx 1.16\rho_w$ and $\rho_2 \approx 1.02\rho_w$
and the coordinates of the layer boundaries are $z_1 \approx 15$\,\AA{} and $z_2 \approx 41$\,\AA. In
the liquid phase $\rho_1 \approx 1.1\rho_w$, $\rho_2 \approx 0.77\rho_w $, $z_1 \approx 18$\,\AA\,and $z_2 \approx 36$\,\AA.

The model profile of the electron density $\rho(z)$ corresponding to Eq. (6) is constructed on the basis of the error function {\bf ${\rm erf}(t)$} [15, 19-23]. The factor $\sigma_{R}$ in the
exponential has the meaning of the standard deviation of the coordinate of the $j$-th boundary in the model
multilayer from its nominal value $z_j$ (see Fig. 1). It takes into account the contribution of capillary waves to the observed structure of the interface and depends on the detector angular resolution. The theoretical value of $\sigma_{R}^2 \approx ( k_BT/2\pi\gamma )\ln(Q_{max}/Q_{min})$  is determined
by the short-wavelength limit in the spectrum of thermal fluctuations of the interface $Q_{max} = 2\pi/a$ (where$a\approx 10$\,{\AA} is about molecular radius) and  the detector angular resolution  $Q_{min}=q^{max}_z\Delta\beta/2$ [16, 24-27]. The calculations of $I_n$ were performed with $q^{max}_z = 0.05$ {\AA}$^{-1}$, whereas $R$ is calculated by Eq. (3) with $q^{max}_z = 0.3$ {\AA}$^{-1}$. Thus, the values of $\sigma_{R}$ lie in the range from 4 to 6\,\AA.

\begin{figure}
\hspace{0.1in}
\epsfig{file=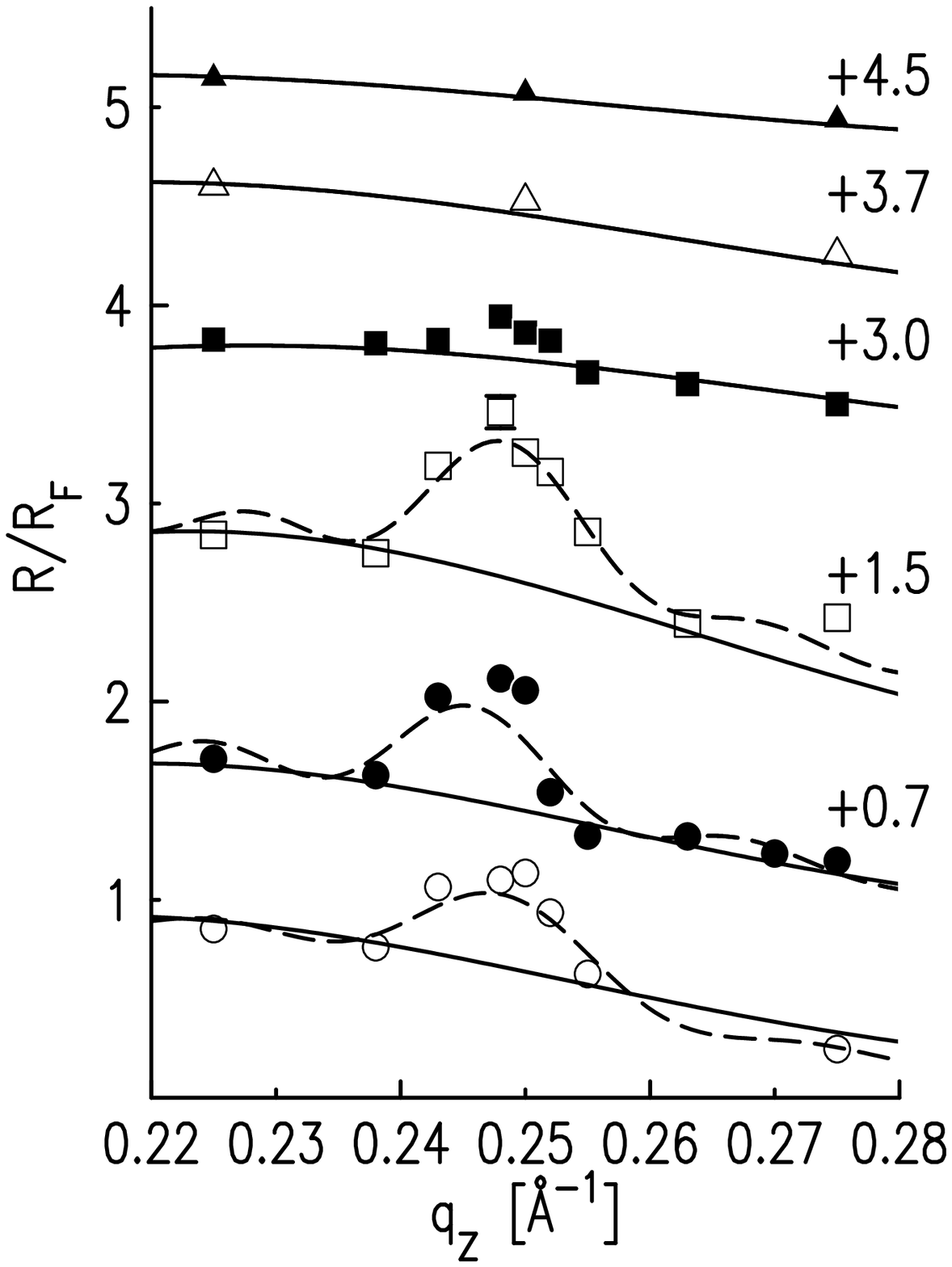, width=0.37\textwidth}

\small {\bf Figure 5}. Peak in the reflection coefficient normalized to $R_F$ at temperatures of 320.6, 322.1, 323.1, 323.5, 324, and 333.4 K (open circles, closed circles, open squares, closed squares, open triangles, and closed triangles, respectively). Solid lines show the results of calculations in the monolayer model (see Eq. (6)). Dashed lines show the results of calculations in the model with the full width of the surface structure of $\sim 340$\,\AA{} ($N=8$) and $W\approx 50$\,\AA{} (see Eq. (8)). The curves are shifted vertically by a value indicated next
to each one for clarity.

\end{figure}

\begin{figure}
\hspace{0.1in}
\epsfig{file=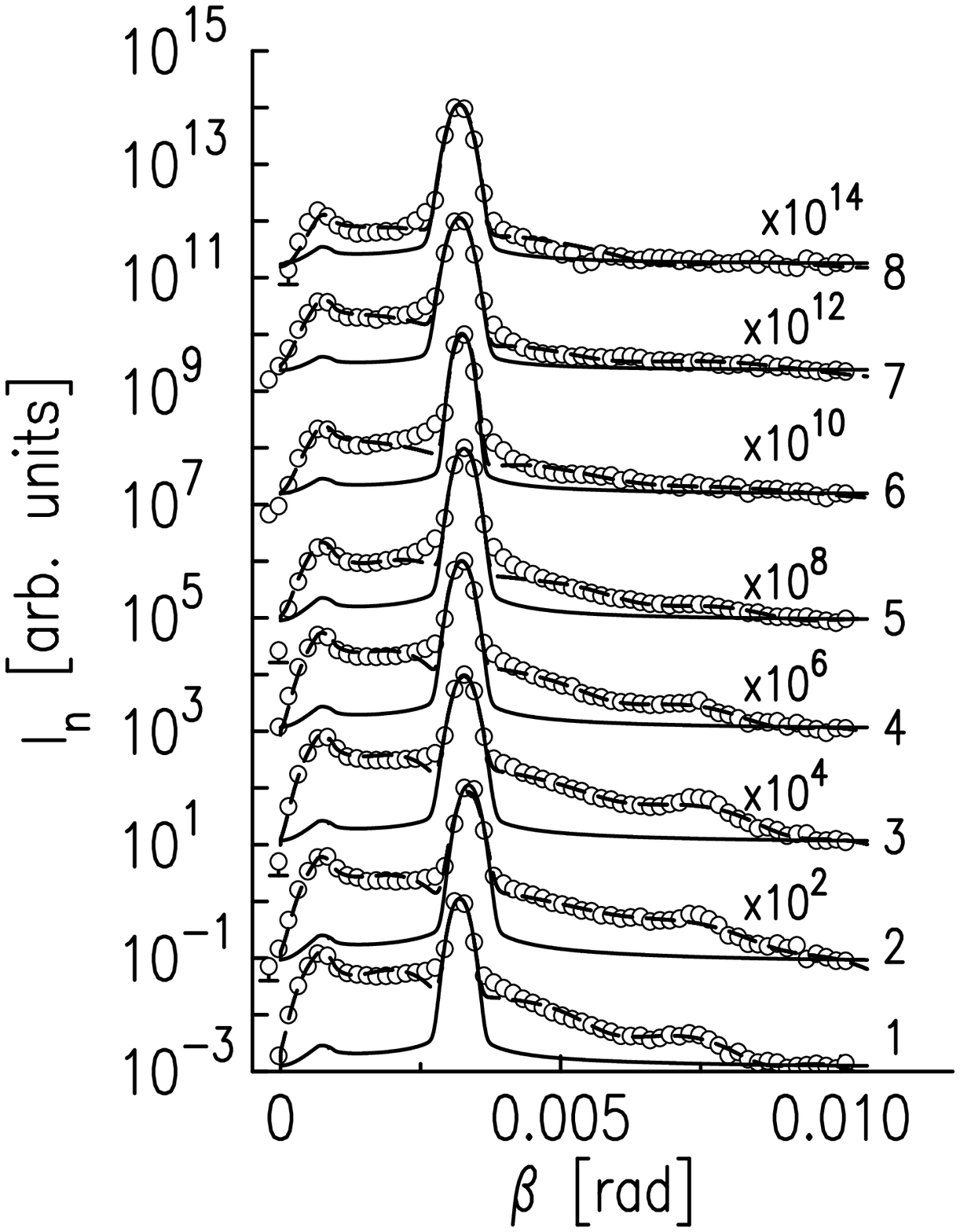, width=0.37\textwidth}

\small {\bf Figure 6}. Intensity of interfacial scattering $I_n$ at the n-hexane~-- water interface for the grazing angle of $\alpha \approx 3.3 \cdot 10^{-3}$\,rad at temperatures $T$ = (1) 319.3, (2) 320.2, (3) 322.1, (4) 323.1, (5) 323.5, (6) 324, (7) 326, and (8) 329.7 K. Solid lines show the results of calculations in the monolayer model (see Eq. (6)). Dashed lines show the results of calculations in the model with an extended layer (Eq. (7) for $T > T_c$ and Eq. (8) with $N=8$ for $T < T_c$).

\end{figure}

The reflection coefficient $R(q_z)$ and scattering intensity $I_n$ calculated using the structure factor $\Phi(q)_m$ are shown in Figs. 4–6 by solid lines. The calculated curves fit the experimental dependences $R(q_z)$ fairly well, and the fitting parameters agree with those from [6] within the error limits. However, the narrow interference peak at $q_z\approx 0.25$\,\AA$^{-1}$, observed for $T < T_c$,
cannot be reproduced, because it corresponds to some structure with a thickness of $2\pi/\delta q_z \sim 300 $\,\AA{}. Furthermore, in the solid phase of the monolayer ($T < T_c$), the observed scattering intensity $I_n$ is about two orders of magnitude higher than the one calculated according
to this model. In the liquid phase ($T > T_c$), the diffuse background intensity decreases with increasing temperature but remains considerably higher than the one predicted by the model using Eq. (6).

The behavior of the system is similar to that described in [5]: the intensity of nonspecular scattering
is nearly independent of the temperature below $T_c$ and decreases gradually with increasing temperature
for $T>T_c$. The reflection coefficient $R$ changes almost stepwise at $T_c$. The low-temperature solid
phase is characterized by an extremely high level of diffuse background, which is $\sim 10^{-1}$ of the specular reflection peak amplitude. The essential difference of the present experimental results from those reported in [5] is the occurrence of peaks in both diffuse scattering and reflection coefficient disappearing upon an increase in the temperature in a narrow range near $T_c$.

The excess scattering intensity observed for $T > T_c$ can be described by the simplest three-layer model
with a structural factor suggested in [5]:
\begin{equation}
\displaystyle
\Phi(q)=\Phi^*_m(q) +\frac{ \displaystyle \delta\rho e^{-\sigma^2q_z^2/2 }}{ \displaystyle \Delta\rho }  e^{-iq_zz_3}.
\end{equation}
Here, the second term describes the uniform third layer with a thickness of $z_3-z_2$ and a density of
$\rho_3=\rho_h + \delta\rho$ (see Fig. 1), the parameter $\sigma$ represents the intrinsic width of the interface between this layer and the bulk of n-hexane, and $\Phi^*_m(q)$ is given by Eq. (6) with the substitution $\rho_3 \to \rho_h + \delta\rho$. The results of calculations according to the three-layer model are shown in Figs. 4–6 by dashed lines. The combined analysis of the data for $I_n$ and $R(q_z)$ demonstrates that the contribution from the second term in Eq. (7) drops rapidly with increasing $q_z$ and becomes negligible for $q_z> 0.075$\,\AA$^{-1}$. The thickness of the third layer is $z_3-z_2 \approx 100 \div 200$\,\AA, the parameter $\delta\rho \approx 0.1 \rho_w \div 0.2 \rho_w$, and the interface width is $\sigma \approx 30 \div 70$\,\AA.

The peaks in diffuse scattering and reflection coefficient observed at $T < T_c$ are described by the structure factor
\begin{equation}
\displaystyle
\Phi(q)=\Phi^*_m(q)+\frac{e^{-\sigma_R^2q^2/2}}{\Delta\rho}\sum_{j=2}^{N}{(\rho_{j+1}-\rho_j)e^{-iqz_j}},
\end{equation}
where the second term describes the planar periodic multilayer (smectic) structure of layer 3 with a period of $W = z_{j+1} - z_j$ (for $j \geq 2$), which determines the positions of the interference maximum on the scattering curves and of the narrow peak in the reflection coefficient. Models with $W=50\pm 5$\,\AA{}, the number of smectic layers $N-2=6 .. 8$, and $\rho_{j} - \rho_h < 0.1 \rho_w$ (for $j \geq 2$) yield satisfactory agreement with the experimental data on both $R(q_z)$ and $I_n(\beta)$ (dashed lines in Figs. 4-6).

Figure 7 shows examples of the electron density profiles $\rho(z)$ for the discussed structures. Structure A, corresponding to $T < T_c$, consists of a solid monolayer with a thickness of $\approx 41$\,\AA{} and a layer with a thickness of $300 \div 400$\,\AA{} and with smectic ordering of the amphiphilic molecules of C$_{30}$-acid. Upon an increase in the temperature, the entire surface structure melts abruptly at $T_c \approx 323.5$\,Ê. Thus, structure B, existing at $T > T_c$, consists of a liquid Gibbs monolayer
with a thickness of $\approx 36$\,\AA{} and a layer of highmolecular-weight alkane liquid with a thickness of $100 \div 200$\,\AA. Unfortunately, the available data are insufficient to reliably establish both the presence of lamellar (bilayer) ordering at $T < T_c$ and the occurrence of orientational (nematic) ordering of the C$_{30}$-acid molecules in layer 3 at $T > T_c$.

Lyotropic and thermotropic mesophases are frequently observed in the bulk and in adsorption films at
interfaces between melts and solutions containing asymmetric amphiphilic molecules [28-32]. For
example, at the surface of high-molecular-weight saturated hydrocarbons and monatomic alcohols (their
interface with air), there occurs a solid–liquid phase transition at a temperature above the bulk melting
temperature [33, 34]. The observation of two-dimensional solid–liquid and liquid–gas phase transitions at
the oil–water interface has also been reported [32, 35-37]. Many authors consider these thermotropic transformations in the context of mono- and bimolecular layer models. The uniqueness of the reversible phase
transition observed in this work is that, upon a decrease in the temperature, a smectic structure is
formed in a $\sim 10$-monolayer-thick surface layer.

Investigations of lyotropic and thermotropic phase transitions between bulk mesophases in aqueous solutions
of fatty acids suggest that one of the parameters that determine the thermodynamic state of the system
is the pH level of the solution, which affects the degree of ionization of the polar groups in amphiphilic molecules [38, 39]. Thus, it is reasonable to associate the dependence $\gamma(T)$ of on the pH level evident in Fig. 2 with the ionization of -COOH in melissic acid. We also note that the enthalpy of the described transition in a system with pH = 10 (partially ionized interface) is a factor of 8 lower than that reported previously for a system with pH = 2 (protonated interface). It is interesting in this context to investigate the behavior of this system in the region of high pH levels ($>12$), where the interface is completely ionized.

\begin{figure}
\hspace{0.1in}
\epsfig{file=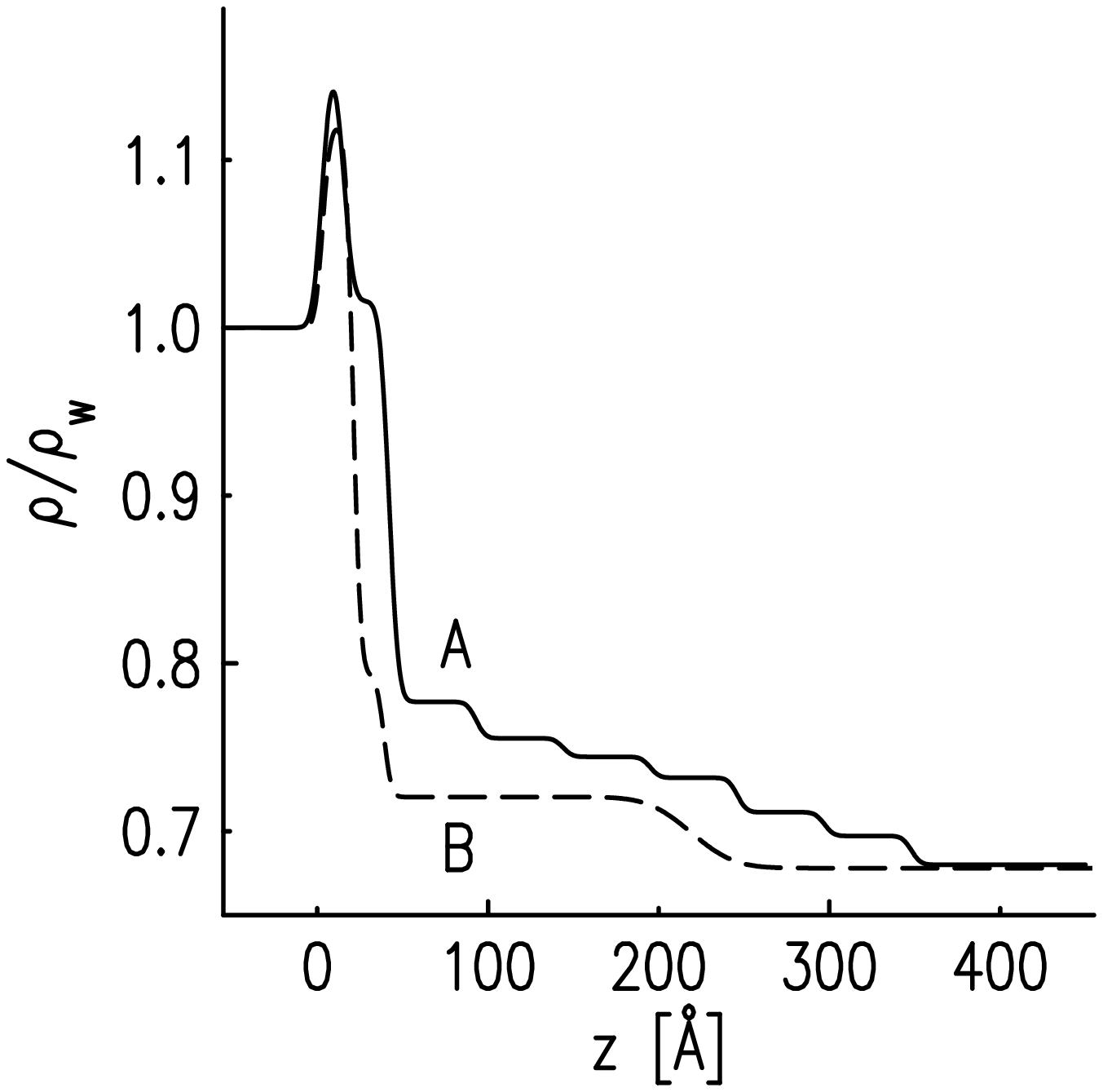, width=0.4\textwidth}

\small
{\bf Figure 7}. Model profiles of the electron density $\rho(z)$ normalized to that in water under normal conditions $\rho_w=0.333$  {\it e$^-$/}{\AA}$^3$: (A) smectic structure and a solid monolayer ($(T<T_c)$, Eq. (8)) and (B) three-layer model with a liquid monolayer ($(T>T_c)$, Eq. (7)).
\end{figure}

This work was performed using the resources of the National Synchrotron Light Source, US Department of Energy (DOE) Office of Science User Facility, operated for the DOE Office of Science by the Brookhaven
National Laboratory under contract no. DE-AC02-98CH10886. The X19C beamline was supported by
the ChemMatCARS National Synchrotron Resource, University of Chicago, University of Illinois at Chicago,
and Stony Brook University.


\small
\begin{thebibliography}{49}
\small
\bibitem{LL5}
L.\,D.\,Landau and E.\,M.\,Lifshitz, {\it Course of Theoretical Physics}, Vol. 5: {\it Statistical Physics} (Nauka, Moscow, 1995; Pergamon, Oxford, 1980).

\bibitem{Takiue}
T.\,Takiue, A.\,Yanata, N.\, Ikeda, K.\,Motomura, M.\,Aratono, J. Phys. Chem.  100, 13743, (1996).

\bibitem{Takiue1}
T.\,Takiue, T.\,Matsuo, N.\,Ikeda, K.\,Motomura, M.\,Aratono, J. Phys. Chem. B 102, 4906 (1998).

\bibitem{CC}
A.\,M.\,Tikhonov, J. Exp. Theor. Phys. 110, 1055 (2010).

\bibitem{acid-c30-3}
A.\,M.\,Tikhonov, JETP Lett. 102, 552 (2015).

\bibitem{acid-c30-2}
A.\,M.\,Tikhonov, JETP Lett. 104, 309 (2016).

\bibitem{s-cell}
D.\,M.\,Mitrinovic, Z.\,J.\,Zhang, S.\,M.\,Williams, Z.\,Q.\,Huang and M.\,L.\,Schlossman, J. Phys. Chem. B 103, 1779 (1999).

\bibitem{acid-c30-1}
A.\,M.\,Tikhonov, H.\,Patel, S.\,Garde, and M.\,L.\,Schlossman, J. Phys. Chem. B 110, 19093 (2006).

\bibitem{alkanes}
A.\,Goebel, K.\,Lunkenheimer, Langmuir 13, 369 (1997).

\bibitem{Adamson}
A.\,W.\,Adamson, {\it Physical Chemistry of Surfaces}, 3rd ed.; John Wiley \& Sons: New York, 1976.

\bibitem{x19c}
M.\,L.\,Schlossman, D.\,Synal, Y.\,Guan, M.\,Meron, G.\,Shea-McCarthy, Z.\,Huang, A.\,Acero, S.\,M.\,Williams, S.\,A.\,Rice, P.\,J.\,Viccaro, Rev. Sci. Instrum. 68, 4372 (1997).

\bibitem{Yoneda}
Y.\,Yoneda, Phys. Rev. 131, 2010 (1963).

\bibitem{Sinha}
S.\,K.\,Sinha, E.\,B.\,Sirota, S.\,Garoff, and H.\,B.\,Stanley, Phys. Rev. B 38, 2297 (1988).

\bibitem{Sinha2}
S.\,K.\,Sinha in {\it Diffuse Scattering and the Fundamental Properties
of Materials}, Edited by R.\,I.\,Barabash, G.\,E.\,Ice, P.\,E.\,A.\,Turchi, Momentum Press, LLS, New Jersey, 2009.

\bibitem{CW}
F.\,P.\,Buff, R.\,A.\,Lovett, F.\,H.\,Stillinger, Phys. Rev. Lett. 15, 621 (1965).

\bibitem{Schwartz}
D.\,K.\,Schwartz, M.\,L.\,Schlossman, E.\,H.\,Kawamoto, G.\,J.\,Kellogg, P.\,S.\,Pershan, B.\,M.\,Ocko,
Phys. Rev. A 41, 5687 (1990).

\bibitem{McClain}
B.\,R.\,McClain, D.\,D.\,Lee, B.\,L.\,Carvalho, S.\,G.\,J.\,Mochrie, S.\,H.\,Chen, J.\,D.\,Litster, Phys. Rev. Lett. 72, 246 (1994).

\bibitem{MWS}
D.\,M.\,Mitrinovic, S.\,M.\,Williams, M.\,L.\,Schlossman, Phys. Rev. E 63, 021601 (2001).

\bibitem{Weeks}
J.\,D.\,Weeks, J. Chem. Phys. 67, 3106 (1977).

\bibitem{Braslau2}
A.\,Braslau, M.\,Deutsch , P.\,S.\,Pershan, A.\,H.\,Weiss, J.\,Als-Nielsen, J.\,Bohr, Phys. Rev. Lett. 54, 114 (1985).

\bibitem{Daillant2}
J.\,Daillant, L.\,Bosio, B.\,Harzallah, and J.\,J.\,Benattar, J. Phys. II 1, 149 (1991).

\bibitem{SCH211}
M.\,L.\,Schlossman, M.\,Li, D.\,M.\,Mitrinovic, A.\,M.\,Tikhonov, High Performance Polymers 12, 551 (2000).

\bibitem{MingLi}
M.\,Li, D.\,J.\,Chaiko, A.\,M.\,Tikhonov, M.\,L.\,Schlossman, Phys. Rev. Lett. 86, 5934 (2001).

\bibitem{Hanley1}
L.\,Hanley, Y.\,Choi, E.\,R.\,Fuoco, F.\,A.\,Akin, M.\,B.\,J.\,Wijesundara, M.\,Li, A.\,M.\,Tikhonov, M.\,L.\,Schlossman, Nucl. Instrum. Methods Phys. Res. B 203, 116 (2003).

\bibitem{Hanley2}
F.\,A.\,Akin, I.\,Jang, M.\,L.\,Schlossman, S.\,B.\,Sinnott, G.\,Zajac, E.\,R.\,Fuoco, M.\,B.\,J.\,Wijesundara, M.\,Li, A.\,M.\,Tikhonov, S.\,V.\, Pingali, A.\,T.\,Wroble, L.\,Hanley
J. Phys. Chem. B 108, 9656 (2004).

\bibitem{Tikh311}
A.\,M.\,Tikhonov, J. Chem. Phys. 124, 164704 (2006).

\bibitem{Tikh111}
A.\,M.\,Tikhonov, J. Phys. Chem. C 111, 930 (2007).

\bibitem{LioLCrys}
P.\,S.\,Pershan, Physics Today 35, 5, 34 (1982).

\bibitem{LioMesophase}
A.\,A.\,Vedenov and E.\,B.\,Levchenko, Sov. Phys. Usp. 26, 747 (1983).

\bibitem{Takiue2}
T.\,Takiue, T.\,Tottori, K.\,Tatsuta, H.\,Matsubara, H.\,Tanida, K.\,Nitta, T.\,Uruga, M.\,Aratono, J. Phys. Chem. B 116, 13739 (2012).

\bibitem{Daillant}
J.\,Daillant, E.\,Bellet-Amalric, A.\,Braslau, T.\,Charitat, G.\,Fragneto, F.\,Graner, S.\,Mora, F.\,Rieutord, and B.\,Stidder, PNAS 102, 11639 (2005).

\bibitem{TAMSCH}
A.\,M.\,Tikhonov, M.\,L.\,Schlossman, J. Phys.: Condens. Matter 19, 375101 (2007).

\bibitem{SurfaceFAlkanes}
B.\,M.\,Ocko, X.\,Z.\,Wu, E.\,B.\,Sirota, S.\,K.\,Sinha, O.\,Gang, and M.\,Deutsch
Phys. Rev. E 55, 3164 (1997).

\bibitem{SurfaceFAlkanols}
O.\,Gang, X.\,Z.\,Wu, B.\,M.\,Ocko, E.\,B.\,Sirota, and M.\,Deutsch
Phys. Rev. E 58, 6086 (1998).

\bibitem{Zhang}
Z.\,Zhang,  D.\,M.\,Mitrinovic, S.\,M.\,Williams ,Z.\,Huang  and M.\,L.\,Schlossman, J. Chem. Phys.
110, 7421 (1999).

\bibitem{LeiBain}
Q.\,Lei, C.\,D.\,Bain, Phys. Rev. Lett. 92, 176103 (2004).

\bibitem{Tamam}
L.\,Tamam, D.\,Pontoni, Z.\,Sapir, Sh.\,Yefet, E.\,Sloutskin, B.\,M.\,Ocko, H.\,Reichert, and M.\,Deutsch, PNAS 108, 5522 (2011).

\bibitem{Cistola}
D.\,P.\,Cistola, D.\,M.\,Small, J.\,A.\,Hamilton, J. Lipid Res. 23, 795 (1982).

\bibitem{Small}
D.\,M.\,Small, {\it The Physical Chemistry of Lipids}, Plenum Press, New York, 1986.


\end{thebibliography}
\end{document}